\documentclass[showpacs,twocolumn,aps,prl,superscriptaddress,final]{revtex4}

\usepackage[dvips]{graphicx}

\begin{document}

\title{Experimental implementation of non-Gaussian attacks \\
on a continuous-variable quantum key distribution system}

\author{J\'er\^ome Lodewyck}
\affiliation{Thales Research and Technologies, RD 128,
 91767 Palaiseau Cedex, France}
\affiliation{Laboratoire Charles Fabry de l'Institut d'Optique, CNRS UMR 8501,\\
Campus Universitaire, b\^at 503, 91403 Orsay Cedex, France}
\author{Thierry Debuisschert}
\affiliation{Thales Research and Technologies, RD 128,
 91767 Palaiseau Cedex, France}
\author{Ra\'ul Garc\'ia-Patr\'on}
\affiliation{QuIC, \'Ecole Polytechnique, CP 165,
Universit\'e Libre de Bruxelles, 1050 Bruxelles, Belgium}
\author{Rosa Tualle-Brouri}
\affiliation{Laboratoire Charles Fabry de l'Institut d'Optique, CNRS UMR 8501,\\
Campus Universitaire, b\^at 503, 91403 Orsay Cedex, France}
\author{Nicolas J. Cerf}
\affiliation{QuIC, \'Ecole Polytechnique, CP 165,
Universit\'e Libre de Bruxelles, 1050 Bruxelles, Belgium}
\author{Philippe Grangier}
\affiliation{Laboratoire Charles Fabry de l'Institut d'Optique, CNRS UMR 8501,\\
Campus Universitaire, b\^at 503, 91403 Orsay Cedex, France}

\begin{abstract}
An intercept-resend attack on a continuous-variable quantum key distribution
protocol is investigated experimentally. By varying the interception fraction,
one can implement a family of attacks where the eavesdropper totally controls
the channel parameters. In general, such attacks add excess noise in the
channel, and may also result in non-Gaussian output distributions. We implement
and characterize the measurements needed to detect these attacks, and evaluate
experimentally the information rates available to the legitimate users and the
eavesdropper. The results are consistent with the optimality of Gaussian attacks
resulting from the security proofs.

\end{abstract}


\maketitle

	Quantum key distribution (QKD) enables two distant parties -- Alice and Bob
-- linked by a quantum channel and an authenticated classical channel to share a
common secret key that is unknown to a potential eavesdropper Eve. For this
purpose, Alice and Bob have to agree on a proper set of non-commuting quantum
variables, as well as a proper encoding of the key into these variables. Common
QKD setups use so-called discrete variables (e.g. the polarization of a photon),
thereby requiring single-photon sources or detectors~\cite{gisin}.

	In this Letter, we shall rather follow an alternative procedure, pioneered
in~\cite{hillery:CVQKD, ralph:CVQKD}, which consists in encoding the key into
continuous variables (CV). Specifically, we use a CVQKD protocol with coherent
states intoroduced in~\cite{fred:nature}. The action of a possible eavesdropper
then appears as {\it added noise} on the observed continuous data. More
precisely, line losses correspond to a restricted class of attacks, often called
beam-splitting attacks, which only add Gaussian ``vacuum" noise. Other attacks
typically add more noise, called ``excess noise", which may be non-Gaussian. It
is generally crucial to show that Alice and Bob can measure these noises with
the required accuracy in order to ensure the security of CV-QKD.

	In order to analyse these noises, we have explicitly implemented several
non-trivial actions of the eavesdropper Eve, which are simple but general
enough to include both Gaussian and non-Gaussian features. These attacks are
implemented optically as partial intercept-resend (IR) operations, in which the
signal beam is either measured and subsequently re-prepared, or is eavesdropped
using a beam splitter (BS). These attacks enable Eve to control independently
the two main channel parameters, namely the loss (BS part) and excess noise (IR
part), simply by adjusting the intercepted fraction. They are therefore much
more powerful than a simple BS attack corresponding to a pure line loss. We
examine in detail how well Alice and Bob can detect them in real operating
conditions. The experiment confirms and emphasizes that it is crucial to
properly evaluate the channel excess noise in order to warrant the security of
the present CV-QKD protocol~\cite{fred:nature,fred:QIC03,hirano}. In addition,
we explicitly measure the information gained by Eve for a wide range of partial
IR attacks, and check that it never exceeds the bound based on Gaussian attacks
(with excess noise). This is in full agreement with the security proof given
in~\cite{fred:ng}.

	Our CV-QKD protocol is based on coherent states and reverse reconciliation,
as described in~\cite{fred:nature}. Alice sends Bob a train of coherent states
$|x+ip\rangle$ where the quadratures $(x,p)$ are randomly chosen from a
bivariate Gaussian distribution with variance $V_A$. Bob randomly measures
either $x$ or $p$, and publicly announces his choice. A binary secret key is
then extracted from the correlated continuous data by using a sliced
reconciliation algorithm~\cite{gilles:slice, bloch:LDPC}. This protocol is well
suited for practical QKD because it only requires conventional fast
telecommunication components, such as InGaAs photodiodes or electro-optics
modulators. A full QKD setup with a typical repetition rate of 1 MHz can be
assembled with off-the-shelves components~\cite{lodewyck:pra}. The security of
the protocol is proven against a wide range of attacks, namely Gaussian
individual attacks~\cite{fred:nature}, finite-size non-Gaussian
attacks~\cite{fred:ng}, and Gaussian collective attacks~\cite{gr,Navascues}. It
will be sufficient for our needs here to focus on the proof of~\cite{fred:ng},
which provides a simple analytical expression for the secret key rates against
non-Gaussian attacks.

\paragraph{General framework.}

	The processing of a coherent state via a Gaussian quantum channel can be
described as follows. Its amplitude is multiplied by $\sqrt{T}$, where $T\leq1$
is the channel transmission, while its noise variance is increased to
$(1+T\epsilon)N_0$ at the output, where $N_0$ stands for the shot-noise level
and $\epsilon$ is the so-called excess noise (referred to the input). Assuming
that the limited efficiency $\eta<1$ of the homodyne detector deteriorates Bob's
reception but does not contribute to Eve's information (so-called ``realistic
mode" in~\cite{fred:nature,lodewyck:pra}), the information rates can be written
as
\begin{eqnarray}
	I_{AB} & = & \frac 1 2 \log_2 \frac{\eta T V_A+1+\eta T \epsilon}{1+\eta T \epsilon}\\
	I_{BE} & = & \frac 1 2 \log_2 \frac{\eta T V_A+1+\eta T \epsilon}
	{\eta/\left[1-T +T \epsilon+ \frac{T}{V_A+1}\right] + 1 - \eta}.
	\label{eq:IBE}
\end{eqnarray}
In reverse reconciliation~\cite{fred:nature}, the secret key rate is given by $
K = \beta \; I_{AB} - I_{BE}$, where $ \beta$ is the efficiency of the
reconciliation algorithm with respect to Shannon's limit. All the quantities
appearing in these formulas are known or can be measured by Alice and Bob. In
practice, Alice and Bob must carefully evaluate $T$ and $\epsilon$ in order to
infer the optimal attack Eve can perform, and therefore to upper bound $I_{BE}$.
This is done by statistical evaluation over a random subset of the raw
data~\cite{lodewyck:pra}.

\paragraph{Non-Gaussian attacks.}

\begin{figure}
\begin{center}
	\includegraphics[width=0.8\columnwidth]{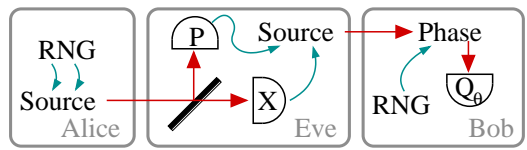}
	\caption{\label{fig:schemaIR} Intercept-resend attack. Alice prepares a
random coherent state, and Bob chooses a random quadrature measurement with a
random number generator (RNG). In between, Eve makes an heterodyne measurement
of each incoming quantum state, and displaces another generated coherent state
according to her measurement result.}
\end{center}
\end{figure}

	Let us consider a particular non-Gaussian attack, namely a partial
intercept-resend (IR) attack: Eve detects and resends a fraction $\mu$ of the
pulses, while she performs a standard beam-splitter (BS) attack on the remaining
fraction $(1-\mu)$. For the IR step, Eve performs a simultaneous measurement of
both quadratures (Fig.~\ref{fig:schemaIR}), and resends a coherent state
displaced according to her measurement results. For the BS step, Eve is assumed
to keep the tapped signal in a quantum memory and to measure it only after Bob
has revealed his measurement basis. For given channel parameters ($T,\epsilon$),
the optimal attack is known to be Gaussian~\cite{fred:ng}. It can be achieved
using an ``entangling cloner"~\cite{fred:QIC03}, which simply reduces to a BS
attack if $\epsilon=0$. If $\epsilon \neq 0$, the partial IR attack that we
consider here is not optimal, although it has several advantages for our
demonstration purposes. First, it gives Eve a very simple way to exploit the
excess noise of the line in order to gain more information; second, it provides
the opportunity to check explicitly the bound on Eve's information
for non-Gaussian attacks, deduced from Alice and Bob's noise variance
measurements as established in~\cite{fred:ng}.

	Let us emphasize that for a full IR attack ($\mu = 1$), one has $\epsilon =
2$. This corresponds to the ``entanglement breaking" limit in our
protocol~\cite{fred:QIC03,nl}, at the edge between the classical and quantum
regimes. No entanglement can be transmitted through the quantum channel, and
therefore no secret key can be extracted. For a lossy channel, this added noise
gets attenuated, so the entanglement breaking limit may become difficult for Bob
to detect. Thus, as another challenge to our experimental implementation, it is
interesting to check whether Bob can detect this IR attack and properly reject
the transmitted key.

\begin{figure}
\begin{center}
	\includegraphics[width=\columnwidth]{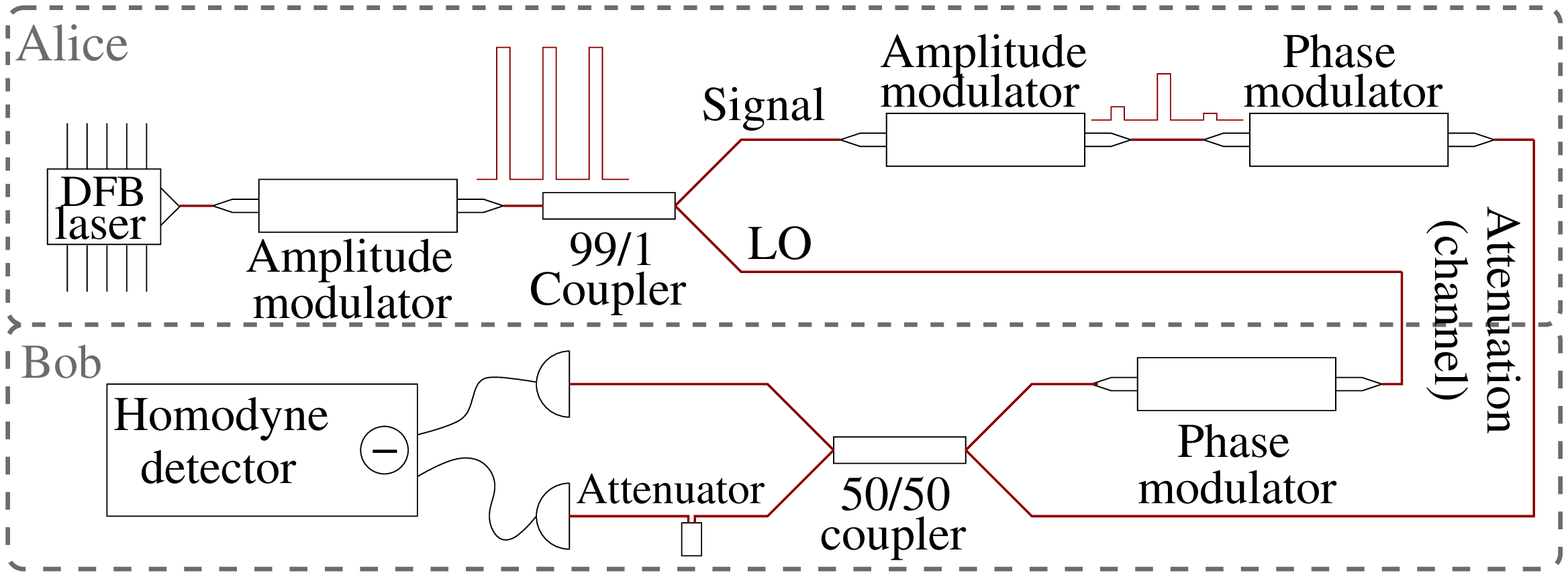}
	\caption{\label{fig:setup}Experimental setup. Alice generates modulated
signal pulses. Bob measures a random quadrature with a pulsed, shot noise
limited homodyne detector.}
\end{center}
\end{figure}

\paragraph{Experimental setup.}

	We have realized the IR attack using the device described
Fig.~\ref{fig:setup}. It is a coherent-state QKD setup, working at 1550 nm and
exclusively assembled with fiber optics and fast telecom components. It
displaces a train of pulsed coherent states within the complex plane, with
arbitrary amplitude and phase, randomly chosen from a two-dimensional Gaussian
distribution with variances $V_A = 36.6\,N_0$. The pulse width is 100 ns. The
signal is sent to Bob along with a strong phase reference -- or local oscillator
(LO), with $10^9$ photons per pulse. Bob selects an arbitrary measurement phase
with a phase modulator placed on the LO path. The selected quadrature is
measured with an all-fiber shot noise limited, time-resolved homodyne detector.
A key transmission is composed of independent blocks of 50000 pulses, sent at a
rate of 500 kHz, among which 10000 test pulses with agreed amplitude and phase
are used to synchronize Alice and Bob and to determine the relative phase
between the signal and LO (see~\cite{lodewyck:pra} for more details). Knowing
this relative phase, Bob is able to choose an absolute phase measurement, for
example one of the field quadratures $x$ and $p$, with a software control loop.

Practical QKD requires that only a part of the dataset is revealed for channel
parameters evaluation. This finite set size introduces statistical fluctuations
that can alter the excess noise estimate.
Therefore, security margins have to be considered when
computing information rates. In all the experimental curves shown below, the
number of sampling points for channel characterization has arbitrarily been
chosen to be 5000 (i.e. 13\% of the 40000 available pulses) for illustration
purpose, and may be optimized for each value of the channel transmission.

\paragraph{Implementation of full IR attacks.}

	To implement an IR attack as in Fig.~\ref{fig:schemaIR}, one would need
three homodyne detectors and two modulation setups. To avoid unnecessary
hardware duplication, this attack has been split in three phases, with the role
of Eve being played either by Alice or by Bob. First, Alice sends coherent
states, and Bob simulates Eve measuring the $x$ quadrature of the incoming
states. Then, the same operation is repeated with a $p$ measurement. To take
into account Eve's beam splitter shown in Fig.~\ref{fig:schemaIR}, the variance
measured by Eve (actually Bob) is adjusted to be exactly half of Alice's output
modulation. This calibration also virtually includes the losses within the
homodyne detector into the beam splitter, thus simulating a perfect heterodyne
measurement. Both $x$ and $p$ measurement outputs are then communicated to Alice
through a classical channel so that she can simulate Eve resending coherent
states that are displaced accordingly. After this sequence, the correlations
between Alice and Bob are measured in order to determine the channel parameters.
Since Alice and Eve drop the quadrature not measured by Bob, our two-step
implementation of the interception is legitimate.


\begin{figure}
\begin{center}
	\includegraphics[width=0.83\columnwidth]{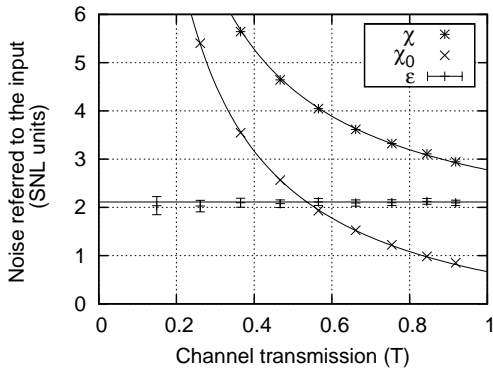}
	\caption{\label{fig:IRnoise} Variance of the noise measured by Bob that is
produced by a full IR attack ($\mu=1$). We define the total added noise
(referred to the input) as $\chi =\chi_0 + \epsilon$, with $\chi_0=1/(\eta T)-1$
denoting the loss-induced vacuum noise and $\epsilon$ denoting the excess noise.
The total added noise $\chi$ (stars) is measured experimentally, while $\chi_0$
(crosses) is deduced from the measured transmission ($T$) and from Bob's
homodyne efficiency ($\eta = 0.6$). Then $\epsilon$ (plus) is obtained from
their difference. The uncertainties margins on $\epsilon$ represent the standard
deviation of statistical fluctuations when computing over a finite data subset
(5000 points out of 40000).}
\end{center}
\end{figure}

	The excess noise referred to the channel input is measured by Bob for
different channel transmissions, selected with an amplitude modulator. For a
full IR attack ($\mu = 1$), the excess noise is measured to be about $0.1\,N_0$
above the expected $2\,N_0$ entanglement breaking bound
(Fig.~\ref{fig:IRnoise}). This is due to the various technical noises
encountered
throughout the IR process, which can be independently determined from the
experimental data (mostly laser phase noise and modulation imperfections).
Since this technical noise is quite small, we also conclude that the
imperfections related to the method used to ``simulate" Eve are negligible.

\paragraph{Implementation of partial IR attacks.}

	Because the full IR attack reaches the entanglement breaking limit, it is
not the best for Eve to tap information from the quantum channel. As explained
previously, a more subtle way for Eve to interact is to intercept and resend
only a fraction $\mu$ of the pulses, and to implement a BS attack on the rest of
the pulses.
In this case, Eve can choose the amount of noise she wants to introduce
independently of the channel transmission. This allows a complete channel
parameter control, which is not achievable with a simple BS attack
($\epsilon=0$) nor with an IR attack where the added noise is fixed for a given
channel transmission.

\begin{figure}
\begin{center}
	\includegraphics[width=0.83\columnwidth]{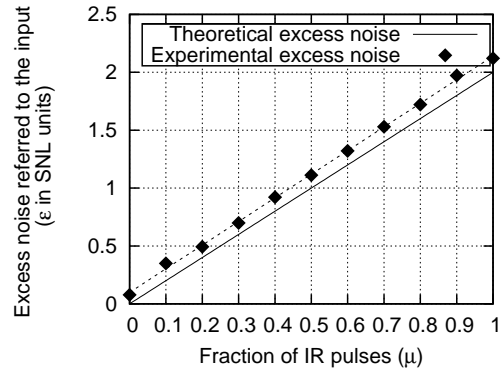}
	\caption{\label{fig:partialIRnoise} Variance of the excess noise in a
partial IR attack. Each point results from an average of the measured excess
noise for different channel transmissions (see the $\epsilon$ vs. $T$ plot of
Fig.~\ref{fig:IRnoise}). The solid line plots the expected excess noise due to
an IR attack on a fraction $\mu$ of the pulses. Due to technical noise, the
experimental data are above this line, typically less than $0.1\,N_0$ (dashed
line).}
\end{center}
\end{figure}

	An important point is that the probability distribution of Bob's
measurements becomes the weighted sum of two Gaussian distributions with
different variances, namely $T V_A + N_0$ for the transmitted data (BS) and
$T(V_A + 2\,N_0) + N_0$ for the resent data (IR), so the attack is not Gaussian
any more. Figure~\ref{fig:partialIRnoise} shows the measured excess noise for
different interception fractions $\mu$. Ideally, it is given by the weighted sum
of the excess noises in the IR and BS cases, \emph{i.e.}, $\epsilon_\mathrm{TOT}
= \mu \; \epsilon_\mathrm{IR} + (1-\mu) \; \epsilon_\mathrm{BS} = 2\mu$. In our
experiment, we have to add the technical noise of variance $\epsilon_\mathrm{T}
= 0.1\,N_0$, which leads to $\epsilon_\mathrm{TOT} = 2\mu +
\epsilon_\mathrm{T}$, in good agreement with the experimental data.

\begin{figure}
\begin{center}
	\includegraphics[width=0.8\columnwidth]{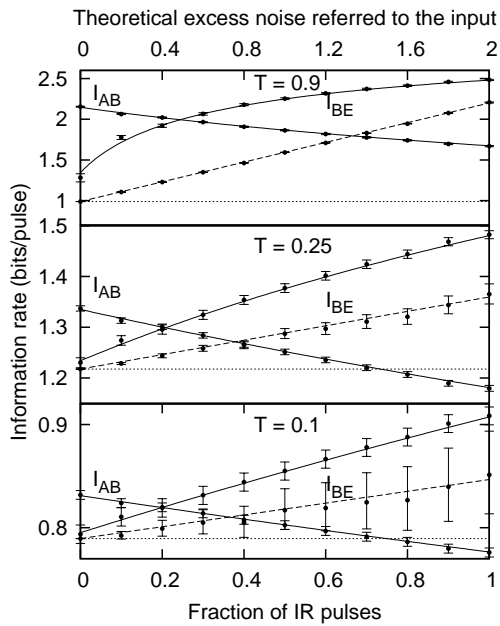}
	\caption{\label{fig:Info} Mutual information rates for a non-Gaussian
partial IR attack, for $T = 0.1$, $0.25$ and $0.9$, with $V_A = 36.6\,N_0$,
$\eta = 0.6$ and a technical excess noise of $0.1\,N_0$. The mutual
information $I_{BE}$ is plotted for a Gaussian model with equivalent excess
noise (solid lines), as well as for a BS attack (dotted lines), and for a
partial IR attack (dashed lines). It is compared with the Gaussian mutual
information $I_{AB}$. As expected, the IR attack enables to exploit the excess
noise, giving Eve extra information above the BS attack. We show statistical
fluctuations ($\pm 1$ standard deviation) corresponding to a data subset of 5000
points per block, as on Fig.~\ref{fig:IRnoise}.}
\end{center}
\end{figure}

	For such an attack, the achievable secret key rate is lower bounded by the
information rate for an equivalent Gaussian attack characterized by the same
variance and conditional variance of the data distribution~\cite{fred:ng}.
The Gaussian mutual information rate $I_{AB}^g$ between Alice and Bob can
be derived from the noise variance measurements with a Gaussian channel model
characterized by the same correlations. This can be compared with the actual
mutual information rate $I_{AB}^{ng}$ computed from the measured data
distribution in presence of the partial IR attack. We find that the Gaussian
mutual information $I_{AB}^g$ is lower than the actual mutual information
$I_{AB}^{ng}$, with a very small gap between them ($\leq 0.8$\% for modulation of
$V_A=36.6\,N_0$) for any $T$ and $\epsilon$. Therefore, only the curve
$I_{AB}^g$ (noted $I_{AB}$) has been represented on Fig.~\ref{fig:Info}.

	On Eve's side, Fig.~\ref{fig:Info} compares $I_{AB}$ with three possible
values of $I_{BE}$. The dotted line ($I_{BE}^{BS}$) is obtained from a BS attack
for the given transmission. With this attack, Eve only makes use of the channel
losses, as if she was not able to exploit the excess noise. The dashed line
($I_{BE}^{partial\ IR}$) is obtained when, in addition to the BS attack, Eve
exploits the excess noise for implementing a partial IR attack. This information
therefore reads
\begin{equation}
	I_{BE}^{partial\ IR} = \mu I_{BE}^{IR} +
	(1-\mu)I_{BE}^{BS},
\end{equation}
The experimental points shown over the dashed line are obtained from this
formula, using the measured information acquired by Eve from the IR part
of the attack $I_{BE}^{IR}$, and the evaluated information from the BS attack
(dotted line). The solid line ($I_{BE}^{g}$) is the optimal Gaussian attack
where Eve exploits the excess noise for implementing an entangling cloner
attack. The experimental points shown over this solid line are the bounds on
$I_{BE}$ deduced from the measured line parameters, according to
eq.~\ref{eq:IBE}. These curves show the crucial role of the excess noise, even
if Eve does not implement the strongest attack.
On Fig.~\ref{fig:Info} one can actually read the tolerable excess noise for
a given channel transmission $T$, at the crossing point between $I_{AB}$ and
$I_{BE}$, confirming that Alice and Bob are
on the ``safe side" when using the Gaussian bound~\cite{fred:ng}.


In conclusion, we have implemented a family of quantum attacks,
namely partial intercept-resend attacks, which allow
Eve to exploit the excess noise and are thus more general than simple
``beam-splitting attacks'' as considered so far. Our experiment confirms
that such attacks can be successfully detected and eliminated by
accurately monitoring the variances of all (``vacuum'' and ``excess'')
noises of the channel. Therefore, the present ``real-case study''
provides both a test and an illustration of the working principles of
experimental CV-QKD. It is also particularly important in view
of the recent proof~\cite{opt1, opt2} that the optimal
{\it collective} attack for a given noise variance is Gaussian, just as
for individual attacks. Considering that our family of attacks
spans all possible relevant transmissions and noise variances of the channel,
the security of our Gaussian-modulated protocol remains warranted
under very general conditions.

We acknowledge financial support from the EU under projects COVAQIAL
(FP6-511004) and SECOQC (IST-2002-506813), and from the IUAP programme of the
Belgian government under grant V-18. R.G-P. acknowledges support from the
Belgian foundation FRIA.

\end{document}